# Comments on "Slowing of Bessel light beam group velocity"

Peeter Saari*

*Institute of Physics, University of Tartu, W. Ostwaldi 1, 50411, Tartu, Estonia*

* *peeter.saari@ut.ee*

**Abstract.** In a recent article [R. R. Alfano and D. A. Nolan, Opt. Commun. 361 (2016) 25] the group velocity reduction below the speed of light in the case of certain Bessel beam pulses has been considered and an idea of its application for a natural optical buffer presented. However, the authors treat the problem as if only one type of Bessel pulse existed, no matter how it is generated. The deficiencies of the article stem from not being familiar with an extensive literature on Bessel pulses, in particular, with a couple of papers published much earlier in the J. Opt. Soc. Am. A, which have studied exactly the same problem more thoroughly.

The Bessel light beams considered in [1] belong to cylindrically symmetric (pseudo) non-diffracting pulsed waves. Since the end of the last century a great variety of such waves has been studied (see the first collective monographs [2, 3]). All of them are different (pulsed) superpositions of monochromatic Bessel modes (beams) and propagate with a subluminal, luminal or superluminal group velocity and without exhibiting diffraction over long distances.

The authors of [1] start from the wave equation in cylindrical coordinates and from the known relationship between the wave numbers of a Bessel mode, viz.,

$$k^2 = k_\perp^2 + k_z^2,$$

where $k = \frac{\omega}{c} n$ (in vacuum $n = 1$) is the wave number and $k_\perp, k_z$ are the transverse and axial wave numbers, respectively. Of course, this relationship—as it is nothing but the wave equation in the Fourier space— always holds no matter what other interdependencies may exist between the variables. But if we look at Eq. (3) of [1], we see that the implicit premise is that $k_\perp$ is an independent constant. In the case of a cylindrical waveguide the value of $k_\perp$ is indeed fixed (for a given mode) by the boundary conditions. But for wave packets in the free 3D space there is no such restriction: $k_\perp$ is a variable which may take any fixed value, run over a range of values independently or act as a function of $k_z$ or the frequency $\omega$, (see review [4] and references therein). The restrictive assumption $k_\perp = const$ which leads to the subluminal group velocity of Bessel wave packets, has been tacitly made earlier as well, the oldest source we know being a handbook [5]. Out of various other possibilities the simplest is the case where both $k_\perp$ and $k_z$ are proportional to the frequency $\omega$ or $k$. In this case the wave packet—called the Bessel-X pulse [6]—not only possesses a *superluminal* group velocity but propagates superluminally as a whole without changing its shape.

As to the Bessel light beam considered in [1], in literature it is named the pulsed Bessel beam and not only its subluminal group velocity but also its whole temporal spread and evolution in the course of propagation have been calculated earlier [7-11].

It follows from the text of [1] that the subluminal Bessel beam can be generated "using an axicon lens or SLM." This statement is misleading. While a circular diffraction grating with constant groove spacing on a SLM indeed corresponds to the condition $k_\perp = const$, an axicon lens or a conical mirror provides the proportionality relation $k_\perp \sim \omega$ and thus generates the Bessel-X pulse which is *superluminal* [2-4, 12, 13]. Propagation and evolution of both the axicon-generated Bessel-X pulse and the grating-generated Bessel pulse have been experimentally investigated [12-15]. Measurements of electric fields in these studies were accomplished with micrometer-range spatial and femtosecond-range temporal resolution and revealed completely the spatio-temporal behavior of the light pulses and thus allowed to determine their group velocities with high accuracy.

The authors of [1] propose to use the subluminal Bessel light beam as a free space delay line or an optical buffer. The same idea—basing on the same equations and formulas—has been thoroughly studied earlier in [16]. In experiments with circular binary phase gratings the subluminality-caused delays up to several hundreds of fs over propagation distance $\sim 10\ cm$ have been recorded [14,15]. Unfortunately, a practical feasibility and usefulness of such an optical buffer which would provide much larger delays (e. g., of the order of hundreds of ps over 1 cm, as considered in [1]) is questionable due to two circumstances. First, due to a group velocity dispersion (GVD)—which is obvious, e. g., from Eq. (14) of [1]—the light pulses broaden and signals become distorted when they are substantially delayed, see the theoretical studies [4, 7-11, 16] and the direct experimental verification in [15]. Second, for a substantial reduction of the group velocity the plane wave components of the Bessel beam must diffract towards the beam propagation axis at large angles close to $90^0$, which means that the diameter of a planar circular grating or a SLM must be much larger than the desired length of the delay line.

In conclusion, the paper [1] promotes study of interesting properties and applications of the pulsed Bessel beams. However, its contents need to be complemented with remarks and references given in this Comment.

This work has been supported by the Estonian Science Foundation (grant PUT369).